\documentclass[nofootinbib,a4paper,aps,pra,showpacs,preprintximbers,twocolumn]{revtex4-2}
\usepackage{cmap}
\usepackage{graphics,graphicx,epsfig}
\usepackage{epstopdf}
\usepackage[centertags]{amsmath}
\usepackage{amsfonts}
\usepackage{amssymb}
\usepackage[utf8]{inputenc}
\usepackage[english]{babel}
\usepackage{graphicx}
\usepackage{amsthm,color}
\usepackage{euscript}
\usepackage{indentfirst}
\usepackage{xcolor}
\usepackage{braket}
\usepackage{type1ec}%
\usepackage{physics}
\usepackage{tabularx}
\usepackage[T2A]{fontenc}	
\DeclareGraphicsExtensions{.pdf,.png,.jpg,.eps}
\begin{document}
	\def\BY{\begin{eqnarray}}
		\def\EY{\end{eqnarray}}
	\def\L{\label}
	\def\nn{\nonumber}
	\def\ds{\displaystyle}
	\def\({\left (}
	\def\){\right )}
	\def\[{\left [}
	\def\]{\right]}
	\def\<{\langle}
	\def\>{\rangle}
	\def\h{\hat}
	\def\td{\tilde}
	\def\r{\vec{r}}
	\def\ro{\vec{\rho}}
	\def\h{\hat}
	\preprint{APS/123-QED}
	
	\title{Entangling two qudits of arbitrary dimension through light-atomic Faraday interaction}
	\author{R. Surmay}%
	\author{V.A. Leonov}%
	\author{E.A. Vashukevich}%

	\affiliation{%
	Saint-Petersburg State University, Universitetskaya Nab. 7/9, St. Petersburg 199034, Russia
	}%

	\begin{abstract}
	   
        The paper investigates entangling operations acting on two qudits of arbitrary dimension, with one encoded in the states of a spatially multimode optical field, the other in those of atomic collective spin coherence, both carrying orbital angular momentum (OAM). We demonstrate the generation of a wide range of entangling operations within a protocol consisting of two Faraday interactions and a rotation of atomic and light quadratures between them. All generated gates can be represented as rational powers of the d-dimensional gate $\hbox{SWAP}_d^\alpha$. The probabilities of two-qudit transformations are calculated for various dimensions of the logical space and for different entangling powers of the logical gates.

	\end{abstract}
		\pacs{42.50.Dv, 42.50.Gy, 42.50.Ct, 32.80.Qk, 03.67.-a}
	\maketitle
	
	\section{\label{Intr} Introduction }

The considerable interest in high-dimensional quantum systems (qudits) can be explained by several reasons. Researchers are attracted by the possibility of increasing the channel capacity, i.e., the amount of information that can be encoded in a single physical carrier. This proves highly useful in quantum communication tasks \cite{Erhard2018a, nguyen2024empowering} and potentially reduces the number of the implemented quantum operations \cite{etxezarreta2023superadditivity}. Significant advantages of qudits over qubits have been demonstrated in quantum cryptography protocols, where security increases with the system dimension \cite{sheridan2010security}. Furthermore, Ref. \cite{muthukrishnan2000multivalued} shows that computations on high-dimensional systems offer advantages over qubit-based systems due to a reduction in the number of required entangling transformations.

Various physical systems for encoding qudits have been proposed. Appropriate systems include time-bin states \cite{bechmann2000quantum, Lukens17}, orbital angular momentum (OAM) of light \cite{Bent2015, li2018atmospheric}, polarization multiphoton states \cite{bogdanov2004qutrit}, transmon systems \cite{krantz2019quantum}, and trapped ions and atoms \cite{burshtein2026robust, hrmo2023native, shi2026efficient}. However, the problem of constructing protocols for highly efficient manipulation of quantum logical states of qudits remains unresolved, as increasing the space dimension introduces additional difficulties in the generation, control, and measurement of high-dimensional quantum states \cite{kiktenko2025colloquium}.

The orbital angular momentum of light is particularly interesting as a resource for qudit computations. Firstly, optical computing is characterized by relatively long state decoherence times \cite{hu2018beating} and, consequently, avoids the problems associated with the simultaneous control of multiple qudit levels. Secondly, OAM can take any integer value, which allows operation in a high-dimensional Hilbert space \cite{Allen92}; moreover, this degree of freedom is natively discrete, thus avoiding the issues caused by the discretization of states in continuous variables. Furthermore, experimental techniques for the generation \cite{slussarenko2011efficient,xiao2016generation, reshetnikov2024quantum}, sorting, and detection of multimode OAM-carrying radiation \cite{mirhosseini2013efficient, leach2002measuring, dai2015measuring} are well developed. Researchers have successfully constructed quantum logic gates on OAM‑based qudits \cite{babazadeh2017high, vashukevich2022high}. However, no methods for implementing entangling transformations for such qudits have yet been proposed \cite{cozzolino2019orbital}. Therefore, the search for physical systems and the development of entangling protocols for states with OAM remains a relevant task.

In Refs. \cite{bashmakova2024parallel,vashukevich2024parallel}, a protocol for entangling two qubits of different physical nature was proposed: one  encoded in the OAM states of a multimode optical field, and the other in the OAM-carrying states of atomic spin coherence. This hybrid system allowed us, firstly, to achieve efficient interaction between the qubits and, secondly, to implement interaction and storage within a single protocol. The atomic qubit was encoded in long-lived states typically employed in memory protocols \cite{moiseev2001complete, moiseev2025optical, muschik2006efficient}. We demonstrated the feasibility of implementing a $\sqrt{\hbox{SWAP}}$ entangling gate with high probability.

The present work generalizes and extends previous work on high-dimensional logical space. We study whether entangling operations between two qudits of arbitrary dimension are feasible within a protocol based on the Faraday (i.e., quantum nondemolition, QND) interaction between multimode OAM-carrying light and an atomic ensemble. The work demonstrates that such a system allows for the definition of qudit states of arbitrary dimension and the generation of a wide class of operations necessary for constructing a universal quantum computer based on high‑dimensional discrete variables. Furthermore, we calculate the probability of implementing entangling transformations as a function of the entangling power and the logical space dimension.
	\section{\label{model} QRQ protocol}
	\subsection{\label{sec:level2}Model}
In the present paper, we consider a model of Faraday (quantum nondemolition, QND) interaction between quantum multimode light with OAM and an ensemble of cold four-level atoms. In Ref. \cite{vashukevich2024parallel} we considered a protocol consisting of two QND interactions separated by a rotation of the atomic and light quadrature operators (hereafter referred to as QND-Rotation-QND, or QRQ). The protocol generates entanglement in the system between two qubits of different physical nature. One qubit resides in the states of a multimode light field, and the other in those of atomic spin coherence. With an appropriate choice of the constants of the two QND interactions and the quadrature rotation angles, such a protocol enables the generation of a broad range of entangling operations, as well as the nonlocal SWAP operation, which is critical for many quantum algorithms \cite{fowler2004implementation}. We now briefly present the basic definitions and approximations of the chosen model.

The light-atomic Faraday interaction in the Holstein--Primakoff approximation can be effectively described by the Hamiltonian \cite{bashmakova2024parallel}
\BY
		\hat{H_I} = & -& \frac{\sqrt{2}\hbar g\sqrt{N}}{\Delta\sqrt{c}}\int dz \sum_{m,l}\left(\chi_{l-m}\Omega_{l-m}[\hat{b}^{\dagger}_{l}\hat{a}_m + \hat{b}_l\hat{a}^{\dagger}_m]\right.\notag \\&- 
		&\left.\chi_{l+m}\Omega_{l+m}[\hat{b}^{\dagger}_l\hat{a}^{\dagger}_m+\hat{b}_l\hat{a}_m]\right), \label{eq:Hamiltonian}   
\EY
where $\hat{a}_j,\hat{b}_j$ are the annihilation operators in the field mode and the atomic spin coherence mode, respectively; the index $j$ denotes the orbital angular momentum; $g$ is the interaction constant, $\Delta$ is the detuning between the quantum field frequency and the atomic transition frequency; $\chi_{l-m}$ is the overlap integral of the transverse spatial profiles of the quantum field mode with index $l$, the spin coherence mode with index $m$, and the driving field mode with index $l-m$, normalized to the beam waist area \cite{vashukevich2022high}; $\Omega_{k}$ is the Rabi frequency associated with a single driving field mode with OAM equal to $k$; and $N$ is the average atomic density. The operators $\hat{b}_l$ are essentially projections of the bosonic spin coherence operator $\hat{b}(\vec r, t)$ between the ground states onto the Laguerre-Gaussian mode spatial profile with index $l$, i.e., annihilation operators in the OAM spin coherence modes.

The resulting Hamiltonian (\ref{eq:Hamiltonian}) consists of two parts: a beam-splitter-type Hamiltonian (the part proportional to $\hat{b}^{\dagger}_{l}\hat{a}_m + \hat{b}_l\hat{a}^{\dagger}_m$) and a parametric-gain Hamiltonian (proportional to $\hat{b}^{\dagger}_l\hat{a}^{\dagger}_m+\hat{b}_l\hat{a}_m$). The presence of both parts in the Hamiltonian ensures that each quantum field mode with index $m$ interacts, via the driving field with index $p$, simultaneously with two atomic modes with indices $m+p$ and $p-m$, and vice versa. To reduce Eq. (\ref{eq:Hamiltonian}) to a Faraday rotation Hamiltonian, we assume that the driving field consists of a single OAM mode with zero angular momentum, i.e., a fundamental Gaussian mode. This value of the OAM of the driving field provides symmetric selective interaction between modes with identical indices in both parts of the Hamiltonian.
The solution of the Heisenberg equations can be written as a Bogoliubov transformation, where the effective coupling constant $\xi$ can be defined as follows:
\BY
	&&	\xi = 2\sqrt{\frac{2N}{c}}g\chi_{0}\sqrt{\int_{0}^{T}\frac{\Omega_{0}^{2}(0,t)}{\Delta^2}}dt.\label{eq:constant_interaction}
\EY
As shown in  Ref. \cite{vashukevich2020} one can reduce all overlap integrals $\chi_l$ for arbitrary $l$ to a single value $\chi_0$ by an appropriate choice of the field geometry. Denoting $\xi_1$ and $\xi_2$ as the constants of the first and second QND interactions, respectively, and  $\theta_1$ and $\theta_2$ as the rotation angles of the light and atomic quadratures between the two interactions, we can write the Bogoliubov transformation as:
\BY
		&&\begin{pmatrix}
			\hat{A}^{\dagger}_m\\
			\hat{B}^{\dagger}_m\\
		\end{pmatrix}^{in} = G \begin{pmatrix}
			\hat{A}^{\dagger}_m\\
			\hat{B}^{\dagger}_m\\
		\end{pmatrix}^{out}  + R\begin{pmatrix}
			\hat{A}_m\\
			\hat{B}_m\\
		\end{pmatrix}^{out}, \label{eq:Bogoliubov} \\
		&&{G} =-\frac{i}{2}\begin{pmatrix}
			2ie^{i\theta_1}+\xi_1\xi_2\sin{\theta_2} & \xi_2e^{i\theta_1}+\xi_1e^{i\theta_2}\\
			\xi_2e^{i\theta_2}+\xi_1e^{i\theta_1} & 2ie^{i\theta_2}+\xi_1\xi_2\sin{\theta_1} \\
		\end{pmatrix},\;\;\;\;\label{eq:Bog_G}\\
	&&	{R} =\frac{i}{2} \begin{pmatrix}
			\xi_1\xi_2\sin{\theta_2} & \xi_2e^{i\theta_1}+\xi_1e^{-i\theta_2} \\
			\xi_2e^{i\theta_2}+\xi_1e^{-i\theta_1} & \xi_1\xi_2\sin{\theta_1}\end{pmatrix}.
	\EY
The operators $\hat{A}_m$, $\hat{B}_m$ in Eq. \eqref{eq:Bogoliubov} are defined as:
\BY
	&&	\hat{A}_m = \frac{\int^T_0{\Omega_0(t)[\hat{a}_m(z,t)+\hat{a}_{-m}(z,t)]dt}}{\sqrt{2\int^T_0{\Omega^2_0(t)dt}}},\label{eq:A_operator}\\
	&&	\hat{B}_m = \frac{1}{2\sqrt{L}}\int^L_0{[\hat{b}_l(z,t)+\hat{b}_{-l}(z,t)]dz},\label{eq:B_operator}
\EY
where $L$ is the length of the atomic ensemble and $T$ is the interaction time. In what follows, we consider the evolution of a two-qubit or, more generally, two-qudit separable input state under the operator transformation given by Eq. (\ref{eq:Bogoliubov}).

	\subsection{\label{sec:citeref}Generation of two-qubit entanglement gates in the QRQ protocol}
Let us first consider the case of interaction between two qubits with logical states defined via the single-excitation physical states of the field and atomic ensemble modes carrying specific OAM. We represent the field qubit state as a single-photon state in a superposition of two modes carrying different OAM values, and the atomic qubit state as a single excitation in a superposition of spin coherence modes with different subscripts. Let the input state be separable:
\BY
		\ket{\psi}_{in} &= &(c_0\ket{0}_1 + c_1\ket{1}_1)\otimes(t_0\ket{0}_2 + t_1\ket{1}_2)\notag\\
		&\equiv& (c^{}_0\hat{A}^{\dagger}_k + c^{}_1\hat{A}^{\dagger}_{k+1})(t^{}_0\hat{B}^{\dagger}_k + t^{}_1\hat{B}^{\dagger}_{k+1})\ket{vac}.\label{eq:qubit_sep}
\EY
Here, the coefficients $c_i$ and $ t_j$ are probability amplitudes normalized as $|c_0|^2+|c_1|^2=|t_0|^2+|t_1|^2=1$. The subscript of the state indicates the qubit number, and the subscript of the operators indicates the mode number. The unnormalized output state after the evolution of the creation operators through the Bogoliubov transformation given by Eq. (\ref{eq:Bogoliubov}) takes the following form:
\BY 
\ket{\psi}_{out}=U_2\ket{\psi}_{in}+ \ket{NQ}+\sigma\ket{vac},
			\label{eq:qubit_out}
\EY
where $\sigma$ is the unnormalized probability amplitude of the vacuum state; $\ket{NQ}$ is a light-atomic state outside the two-qubit space, containing terms with excitation bunching in either the light or the atomic subsystems, $U_2$ is the two-qubit transformation matrix in the QRQ protocol. The subscript 2 denotes the dimension of the logical states under consideration (qubits). The state with excitation bunching can be explicitly written as (here and below, we omit the $out$ subscript on the operators):
	\BY
	\ket{NQ}&\equiv&\left[G_{11}G_{12}(c^{}_0\hat{A}^{\dagger}_0 + c^{}_1\hat{A}^{\dagger}_1)(t^{}_0\hat{A}^{\dagger}_0 + t^{}_1\hat{A}^{\dagger}_1)\right.\nn\\
&+&\left.G_{21}G_{22}(c^{}_0\hat{B}^{\dagger}_0 + c^{}_1\hat{B}^{\dagger}_1)(t^{}_0\hat{B}^{\dagger}_0 + t^{}_1\hat{B}^{\dagger}_1)\right]\ket{vac}.\;\;\;\;\;\;
	\EY
In the expression above, $G_{ik}$ are the matrix elements of Eq. (\ref{eq:Bog_G}). We write the state in unnormalized form and will explicitly renormalize it when generalizing the system evolution to qudits. Here we focus on the two-qubit transformation matrix and its properties.

Using the explicit form of the Bogoliubov transformation of Eq. (\ref{eq:Bogoliubov}), one can notice that the matrix $U_2$, which is the implemented logical gate, has only two types of nonzero matrix elements, $\mu$ and $\nu$:
	\BY
	U_2 =\begin{pmatrix}
		\mu+\nu&0&0&0\\
		0&\mu&\nu&0\\
		0&\nu&\mu&0\\
		0&0&0&\mu+\nu\\
	\end{pmatrix}. \label{eq:Udef}
	\EY
Here we introduce notation for the interaction control parameters in terms of the Bogoliubov matrix elements:
		\BY
		\mu\equiv G_{11}G_{22}, \nu\equiv G_{12}G_{21}.\label{munu}
		\EY
Since we consider the evolution of a two-qubit system, we require $U_2$ to be unitary. We impose the condition $U^{\dagger}U = \eta I_2$ on the parameters $\mu$ and $ \nu$, where $I_2$ is the identity matrix in the two-qubit space, and $\eta$ is a real coefficient such that $\sqrt{\eta}$ is the unnormalized probability amplitude for implementing the logical gate under consideration. Writing $U^{\dagger}U$ explicitly, one obtains the following equation for the parameters , along with the expression for the normalization coefficient $\eta$:
\BY
	&&\nu\mu^* + \mu\nu^* = 2(\Re[\nu]\Re[\mu]+\Im[\nu]\Im[\mu])=0,\label{eq:unitar}\;\;\;\;\;\\
	&&\eta=|\mu|^2 + |\nu|^2 \label{eta}.
\EY
From Eqs. (\ref{eq:Bog_G}) and (\ref{munu}), it can be concluded that after factorizing a common phase $e^{i (\theta_1 + \theta_2)}$ from the matrix $U_2$, the matrix element $\nu$ becomes purely real, and the condition given by Eq. (\ref{eq:unitar}) simplifies significantly. In this case, the choice $\Re[\nu]=0$ leads to a trivial identity two-qubit transformation, so the only nontrivial solution of Eq. (\ref{eq:unitar}) is a purely imaginary element $\mu$. The renormalized matrix $\widetilde{U}_2$ (hereafter, renormalized quantities will be denoted by a tilde) can then be written as:
	\BY
&&	\widetilde{U}_2 =\frac{1}{\sqrt{\eta}}U_2=\begin{pmatrix}
		\widetilde{\mu}+\widetilde{\nu}&0&0&0\\
		0&	\widetilde{\mu}&	\widetilde{\nu}&0\\
		0&	\widetilde{\nu}&	\widetilde{\mu}&0\\
		0&0&0&	\widetilde{\mu}+	\widetilde{\nu}\\
	\end{pmatrix}.\label{eq:Urenorm}
	\EY
We also note that $\widetilde{U}_2$ can be represented as a superposition of the identity and SWAP transformations:
	\BY
&&	\widetilde{U}_2 = \widetilde{\mu} I_2+  \widetilde{\nu} \hbox{SWAP}.\label{U2}
\EY
It is precisely the values of the control parameters that determine both the type of the implemented transformation and its success probability. In the next section, we will generalize the approach presented in this part of the work to higher-dimensional logical systems, i.e., qudits. We will analyze in detail the entangling power of the resulting two-qudit transformation and, using the Schmidt decomposition, express it in terms of the control parameters for an arbitrary dimension of the logical space.

	\section{Entangling of two qudits of arbitrary dimension}
		\subsection{Definition of vectors of logical space and normalization of output state}
		
In Section IIB, we described the method for identifying the basis states of the logical two-qubit space via the physical states of various field and atomic modes with single excitations. We can generalize this method to the case of a logical space of dimension $d$. By proceeding in a similar manner, we can define two-qudit systems. For the logical state $\ket{0}$ of the first qudit, we choose the physical field state with a single photon in a mode with an arbitrary OAM index $k$. Similarly, for the $j$-th mode, we define the logical state $\ket{j}$ through the field mode state with index $k+j$. We can choose the states of the second qudit likewise, using the collective spin coherence modes of the atomic ensemble:
\BY
		\ket{j}^{Logic}_1\equiv \h A^\dag_{k+j}\ket{vac}=\ket{1}_{k+j,L}, j \in \{ 0,...,d-1\}, \label{eq:logic_coding_light}\\
		\ket{j}^{Logic}_2\equiv \h  B^\dag_{k+j}\ket{vac}=\ket{1}_{k+j,A}, j \in \{ 0,...,d-1\}. \label{eq:logic_coding_atom}
\EY
Here, $\ket{1}_{k,L}$ and $\ket{1}_{k,A}$ denote the Fock states of the field and atomic modes carrying OAM, respectively. Thus, each two-qudit system contains two excitations: one in the superposition state of $d$ light modes, and another in the superposition state of $d$ spin coherence modes.

The Bogoliubov transformation, given by Eq. (\ref{eq:Bogoliubov}) couples the atomic and field modes identically, irrespective of the OAM index. Therefore, we can describe the evolution of the input two-qudit state similarly to Eq. (\ref{eq:qubit_out}). As the input state, we choose a separable state of two qudits of equal dimension as the input state:
\BY
\ket{\psi}_{in} = 
& \(\sum\limits_{i=0}^{d-1}{c^{}_i\hat{A}^{\dagger}_{k+i}}\)\otimes\(\sum\limits_{j=0}^{d-1}{t^{}_j\hat{B}^{\dagger}_{k+j}}\)\ket{vac},\label{eq:qudit_sep}
\EY
where the sum of the squared amplitudes $c_i, t_i$ for both qudits is normalized to unity. The output state after the system evolves under the Hamiltonian in Eq. \eqref{eq:Hamiltonian} can be written as:
\BY
		\ket{\psi}_{out} &=&U_d\ket{\psi}_{in} +\ket{NQ}_d+\sigma_d\ket{vac},\label{eq:output_state}
\EY
where the notation from Eq. (\ref{eq:qubit_out}) is retained, up to an additional index $d$ indicating the dimension of the logical space.
Then, the state containing all terms with excitation bunching can be written in the following form:
\BY
	\ket{NQ}_d&\equiv&\left[G_{11}G_{12} \sum\limits_{i,j=0}^{d-1}c_i t_jA^\dag_iA^\dag_j \right.\nn\\
&+&\left.G_{21}G_{22} \sum\limits_{i,j=0}^{d-1}c_i t_jB^\dag_iB^\dag_j\right]\ket{vac}.
\EY
This state contains two contributions with excitations grouped in two different media: the light and the atomic ones. Naturally, if we want to track the probability of either contribution, we should renormalize each of them separately. However, in the present work, we are interested only in the probability of bunching as such. Let us then restore the correct normalization of the state:
\BY
&&{}_d\langle NQ | NQ \rangle_d=|G_{11} G_{12}|^2\(\sum_{i,j}|c_i|^2|t_j|^2 +\sum_{i\neq j}c_it_jc^*_jt^*_i\)\notag\\&&+|G_{21} G_{22}|^2\(\sum_{i,j}|c_i|^2|t_j|^2 +\sum_{i\neq j}c_it_jc^*_jt^*_i\)=\notag\\&&\(1 +\sum_{i\neq j}c_it_jc^*_jt^*_i\)\(|G_{11} G_{12}|^2+|G_{21} G_{22}|^2\)\equiv|\tau_d|^2. \label{norm}
\EY
All summations are carried out from 0 to $d-1$. The renormalized state with excitation bunching can then be written as:
\BY
	\widetilde{\ket{NQ}}_d&\equiv&\frac{1}{|\tau_d|}\ket{NQ}.\label{eq:NQrenorm}
\EY
Proceeding in the same manner, one can calculate the probability amplitude for the vacuum state $\sigma_d$:
\BY
\sigma_d\ket{vac}&=&\sum\limits_{i=0}^{d-1} c_i t_i\(\hat{B}^{}_i\hat{B}^{\dagger}_i R_{12}G_{22}+ \hat{A}^{}_i\hat{A}^{\dagger}_i R_{11}G_{21}\)\ket{vac} \notag\\&=&\sum\limits_{i= 0}^{d-1} c_i t_i\( R_{12}G_{22}+ R_{11}G_{21}\)\ket{vac}.\label{eq:Vrenorm}
\EY
Referring to Eqs. (\ref{eq:Urenorm}), (\ref{eq:NQrenorm}) and (\ref{eq:Vrenorm}), we can write the renormalized output state, denoting, as before, the renormalized quantities with a tilde:
 \BY
 \widetilde{\ket{\psi}}_{out} &=&\frac{\sqrt{\eta} \widetilde{U_d}\ket{\psi}_{in} +|\tau_d|\widetilde{\ket{NQ}}_d+\sigma_d\ket{vac}}{\sqrt{|\eta|+|\tau_d|^2+|\sigma_d|^2}}.\label{eq:output_state_norm}
 \EY
The coefficient $\eta$ in the expression above is defined by Eq. (\ref{eta}) and is invariant under the changes in the dimension of the logical space. However, as can be seen from Eqs. (\ref{norm}, \ref{eq:Vrenorm}), the other two probability amplitudes depend not only on the Bogoliubov matrix elements  but also on the coefficients determining the input qudit states. Thus, to proceed with the quantitative analysis, we need to estimate the amplitudes $\tau_d,\sigma_d$ independently of the input state. Since we are interested in the lower bound on the probability of the two-qudit transformation, we determine the maxima of the squared moduli of the probability amplitudes of the vacuum term and the excitation-bunching term. As shown in Appendix A, the maxima of the aforementioned expressions are achieved simultaneously when all qudit probability amplitudes are equal, which, together with the normalization condition, yields $|c_i|  = |t_j| =  \frac{1}{\sqrt{d}}, \{i,j\}\in\[0,d-1\]$.
We call an input state with such probability amplitude values an equal-weight state. Under this condition, we can write the maxima of probability amplitudes used to establish the lower bound on the two-qudit operation probability in the following form:
\BY
&&|\tau^{est}_d|^2=(2-\frac{1}{d})\(|G_{11} G_{12}|^2+|G_{21} G_{22}|^2\),\\
&&|\sigma^{est}_d|^2=|\( R_{12}G_{22}+ R_{11}G_{21}\)|^2.
\EY
The obtained expressions will be used further to find the maximum entangling-transformation probability in the QRQ protocol.

	\subsection{The power of entanglement of two-qudit gates}

As in the two-qubit case, to describe the two-qudit unitary transformation $U_d$, it is convenient to use the same variables $\mu$ and $\nu$. We represent $U_d$ as a decomposition into elementary gates, since the structure of the matrix $U_d$ described by Eq. (\ref{U2}) is preserved as the dimension $d$ changes. The multidimensional logical operations $\hbox{SWAP}_d$ and $I_d$ can be expressed in terms of projectors onto logical states in the computational basis $P_{n,m}:=\ket{n}\bra{m}$ as follows:
\BY
	&&\hbox{SWAP}_d = \sum_{i,j=0}^{d-1}P_{i,j}\otimes P_{j,i}, \label{eq:SWAP}\\
&&	I_d = \sum_{i=0}^{d-1}P_{i,i}\otimes P_{i,i} .\label{eq:I}
\EY
	
We take into account the normalization factor obtained earlier for the matrix $U_2$ from unitarity considerations. Since this factor retains its form for an arbitrary dimension $d$, we can obtain the following expression:
\BY
		\widetilde{U}_{d} =\widetilde{\mu}I_{d} + \widetilde{\nu}\hbox{SWAP}_{d},\label{U_gate}
\EY
	or 
\BY
						\widetilde{U}_{d}  &=&\widetilde{\mu}\sum_{i=0}^{d-1}P_{i,i}\otimes P_{i,i} + \widetilde{\nu}\sum_{i,j=0}^{d-1}P_{i,j}\otimes P_{j,i} \nn\\
			&=&(\widetilde{\mu}+\widetilde{\nu})\sum_{i=0}^{d-1}P_{i,i}\otimes P_{i,i} + \widetilde{\nu}\sum_{i\neq j}^{d-1}P_{i,j}\otimes P_{j,i}.\label{eq:pro}
\EY
A useful tool for further description of two-qudit unitary transformations is the operator Schmidt decomposition. An arbitrary linear operator $U$ can be represented as:
\BY
		U =\sum_n s_n A_n \otimes B_n,
\EY
where $s_n$ are the decomposition coefficients, called Schmidt numbers, and $A_n$, $B_n$ are sets of operators forming a basis of the operator space on systems 1 and 2, respectively. The Schmidt numbers allow one to calculate the linear entropy:
\BY
		E\(U\) = 1 -\frac{1}{d^4}\sum{s^4_n},\label{entropy}
\EY 
where $d$ is the dimension of the system.

In Ref. \cite{wang2003entangling}, it was proposed to use linear entropy to quantify the entanglement of qudit transformations, and it was shown that this measure, called entangling power, satisfies the criteria required in Ref. \cite{horodecki2000limits}. The entangling power $e_p$ for the generated unitary operator $\widetilde{U}_d$ takes the form:	
\BY
		e_p(\widetilde{U}_{d}) = &&\left(\frac{d}{d+1}\right)^{2} \left[E(\widetilde{U}_{d})+E(\widetilde{U}_{d} \hbox{SWAP}_{d})\right.\nn\\
		&&-E(\hbox{SWAP}_{d})\Big{]}.
\EY
Using the Schmidt decomposition for the two-qudit interaction matrix $\widetilde{U}_{d}$, similarly to the method demonstrated in Ref. \cite{ekert1995entangled}, we choose the multidimensional generalization of rotation operators (qubit Pauli operators), termed Heisenberg–Weyl operators, as the operator basis for the Schmidt decomposition in Eq. (32):
	\BY
	 D_{s,l}=\frac{1}{\sqrt{d}}\sum\limits_{k=0}^{d-1} \exp{\frac{2 i \pi s}{d} k}\ket{k\oplus_d l}\bra{k}.
	\EY
Thus, for example, the operator $D_{0,0}$ is simply the identity matrix renormalized by the square root of the space dimension. Since the operation under study $\widetilde{U}_d$ is a superposition of the identity matrix and the $\hbox{SWAP}_d$ matrix, using the explicit form of the Heisenberg-Weyl operators and Eq. (\ref{U_gate}), we can write $\widetilde{U}_d$ as:
\BY
		\widetilde{U}_{d} = d\;\widetilde{\mu}D_{0,0}\otimes D_{0,0}+\widetilde{\nu}(\sum\limits_{l,s=0}^{d-1}D_{l,s}\otimes D_{l,s}),
\EY
whence it is evident that the Schmidt number $s_1 = d\;\widetilde{\mu}+\widetilde{\nu}$, and the remaining $d^2-1$ Schmidt numbers are equal to $\widetilde{\nu}$. Then, the entropy of the unitary transformation $\widetilde{U}_d$ is
\BY
		E(\widetilde{U}_{d}) = 1 - \frac{1}{d^4}\left(|d\;\widetilde{\mu}+\widetilde{\nu}|^4 + |\widetilde{\nu}|^4(d^2 -1)\right).
\EY
To determine the entropy of the $\widetilde{U}_d\hbox{SWAP}_d$ gate, it suffices to write it explicitly using the representation  given by Eq. \eqref{U_gate} for the matrix $U_d$:
\BY
		\widetilde{U}_d\hbox{SWAP}_d &=& (\widetilde{\mu}I_{d} + \widetilde{\nu}\hbox{SWAP}_{d})\hbox{SWAP}_{d}\nn\\
		&=&  \widetilde{\mu}\hbox{SWAP}_{d} + \widetilde{\nu}I_{d}.\label{pffr}
\EY
	
From Eq. \eqref{pffr}, it can be noted that the Schmidt numbers of the $\widetilde{U}_d\hbox{SWAP}_d$ gate coincide with those for $\widetilde{U}_d$ up to the interchange of the coefficients $\widetilde{\mu}$ and $\widetilde{\nu}$. The entropy of such a gate then takes the value:
\BY
		E(\widetilde{U}_{d}\hbox{SWAP}_d) = 1 - \frac{1}{d^4}\(|d\;\widetilde{\nu}+\widetilde{\mu}|^4 + |\widetilde{\mu}|^4(d^2 -1)\).\;\;\;\;
\EY
For the $\hbox{SWAP}_d$ gate, the entropy value is determined as for the $\widetilde{U}_d$ gate at $\widetilde{\nu}=0$:
\BY
		E(\hbox{SWAP}_{d}) = 1 - \frac{1}{d^2}.
\EY
	
Then, the entangling power for an arbitrary gate implemented in the considered protocol, taking into account the unitarity conditions \eqref{eq:unitar}, i.e., assuming $\widetilde{\mu}$ to be purely imaginary and $\widetilde{\nu}$ to be purely real, takes the form:
\BY
		e_p(\widetilde{U}_{d}) = \frac{2(d-1)}{(d+1)}\widetilde{\nu}^2\;\widetilde{\mu}^2.\label{eq:ep_U}
\EY
It can be seen that the entangling power $e_p$ depends both on the interaction parameters and on the dimension of the system, and this dependence factorizes. In the next section, we will identify the class of the performed transformation and show that the conditions for implementing a logical operation with a fixed entangling power do not depend on the dimension of the logical space.
	
	\subsection{Gate $\hbox{SWAP}^{\alpha}_d$}

According to the calculations presented in the previous section, Eq. \eqref{eq:ep_U} shows that the unitary transformation $\widetilde{U}_d$ possesses nonzero entangling power $e_p\neq 0$ for parameter values $\widetilde{\mu}, \widetilde{\nu}\neq 0$. For a detailed description of the $\widetilde{U}_d$ action mechanism, we consider the $\hbox{SWAP}_d^{\alpha}$ gate. We can regard this gate as a power of the $\hbox{SWAP}_d$ gate and define it as follows:
\BY
		\hbox{SWAP}^{\alpha}_{d}=\frac{1}{2}(1+e^{i \alpha \pi}) I_{d} + \frac{1}{2}(1-e^{i \alpha \pi}) \hbox{SWAP}_{d}.\label{def:swap}
\EY

It is straightforward to verify that potentiating the right-hand side of Eq. \eqref{def:swap} to the power $1/\alpha$ yields the $\hbox{SWAP}_d$ gate satisfying Eq. \eqref{eq:SWAP}. Note that definition, given by Eq. \eqref{def:swap} formally resembles Eq. (\ref{U_gate}).

We define the entangling power of the $\hbox{SWAP}_d^{\alpha}$ gate according to Eq. \eqref{def:swap}:
\BY
		e_p(\hbox{SWAP}_d^{\alpha}) = \frac{1}{4} \frac{2(d-1)}{(d+1)}\sin^2{(\alpha\pi)}.\label{eq:ep_swap}
\EY

The maximal entanglement is achieved at $\alpha=(2k+1)/2, k\in\mathbb{Z}$. This value corresponds to the gate $\sqrt{\hbox{SWAP}_d}$ multiplied by the additional factor $\hbox{SWAP}_d {k}$ (an integer value of $k$ does not change the entangling power). In the obtained expression, the factorization  is also noteworthy. One factor depends only on the power $\alpha$ and the other is determined solely by the system dimension. The factors dependent on the system dimension in Eqs. (\ref{eq:ep_swap}) and (\ref{eq:ep_U}) coincide. Hence, the control parameters of the protocol can be related to the entangling power of the gate:
\BY
	\widetilde{\nu}^2\;\widetilde{\mu}^2= \frac{1}{4}\sin^2(\alpha \pi). \label{EnPC}
\EY
From the obtained relation, the choice of the power $\alpha$ provides a second condition on the parameters $\widetilde{\nu}$ and $\widetilde{\mu}$. Taking  Eq. (\ref{eq:unitar}) into account, we then optimize the values of the two parameters to maximize the generation probability of the entangling interaction with a fixed entangling power specified by $\alpha$. Notably, the interaction parameters required to generate the gate are independent of the system dimension and determined solely by $\alpha$. For example, the maximally entangling transformation $\sqrt{\hbox{SWAP}}$ will be generated for both qubits and qudits of arbitrary dimension using the same set of physical parameter values.
				
\section{Probability of generating a logical two-qudit transformation with a fixed entanglement power}
                
The probability of generating the entangling unitary transformation $\widetilde{U}_d$ can be written using Eq. (\ref{eq:output_state_norm}):	\BY
	P_U=\frac{\eta}{|\eta|+|\tau_d|^2+|\sigma_d|^2} .\label{eq:prob}
		\EY
After substituting the worst-case estimates of the amplitudes in Eqs. (26) and (27) into the previous expression, the lower bound on the probability of the unitary transformation is expressed in terms of the Bogoliubov matrix elements given by (\ref{eq:Bog_G}) and (5) and takes the following form:
		\BY
		P^{est}_U&=&\(|G_{11}G_{22}|^2+|G_{12}G_{21}|^2\)\nn\\
		&&\times\left[|G_{11}G_{22}|^2+|G_{12}G_{21}|^2 \right.\nn\\
		&&\left.+(2-1/d)\(|G_{11} G_{12}|^2+|G_{21} G_{22}|^2\)\right.\nn\\
		&&\left.+|\( R_{12}G_{22}+ R_{11}G_{21}\)|^2\right]^{-1}.
		\EY
The matrix elements depend on four physical parameters in the general case. These parameters are the effective coupling constants for the first and second Faraday interactions, $\xi_1$ and $\xi_2$, and two quadrature rotation angles of the light and atomic operators between the interactions, $\theta_1$ and $\theta_2$. We impose two constraints: the unitarity condition of the two-qudit transformation matrix in Eq. (\ref{eq:unitar}) and the fixing of the entangling power in Eq. (\ref{EnPC}). Appendix B provides the analytical solutions to these equations and discusses the number of free parameters.
		
Numerical analysis shows that the highest probability values are achieved when the difference between the rotation angles of the atomic and light qudits satisfies $\theta_1-\theta_2=\pi k, k\in\mathbb{Z}$. In this case, we fix the sum of the angles $\theta_1+\theta_2$ to ensure the unitarity of the transformation given by (\ref{eq:unitar}). We also relate the sum of squares of the effective coupling constants $\xi_1^2+\xi_2^2$ to the entanglement parameter $m=\frac{1}{2}\sin(\alpha \pi)$. We can then construct the probability as a function the single remaining free parameter $q=\xi_1\xi_2 \cos(\theta_1-\theta_2)$. It should be noted that there is some ambiguity in the choice of angles and $\xi_1^2+\xi_2^2$, also discussed in Appendix B. In particular, the probability exhibits branching into two physically distinct operating regimes of the protocol: $\xi_1>\xi_2$ and $\xi_1<\xi_2$. We are interested in the lower bound on the maximum probability, i.e., the maximum probability given the largest contribution of the vacuum and excitation-bunching terms. We therefore consider the behavior of the probabilities for a theoretically maximal qudit-state dimension $d\xrightarrow{}\infty$, for which the dimension-dependent factor in Eq. (46) attains its maximum value.
		
		\begin{figure}[h!]
			\includegraphics[width=8.6cm]{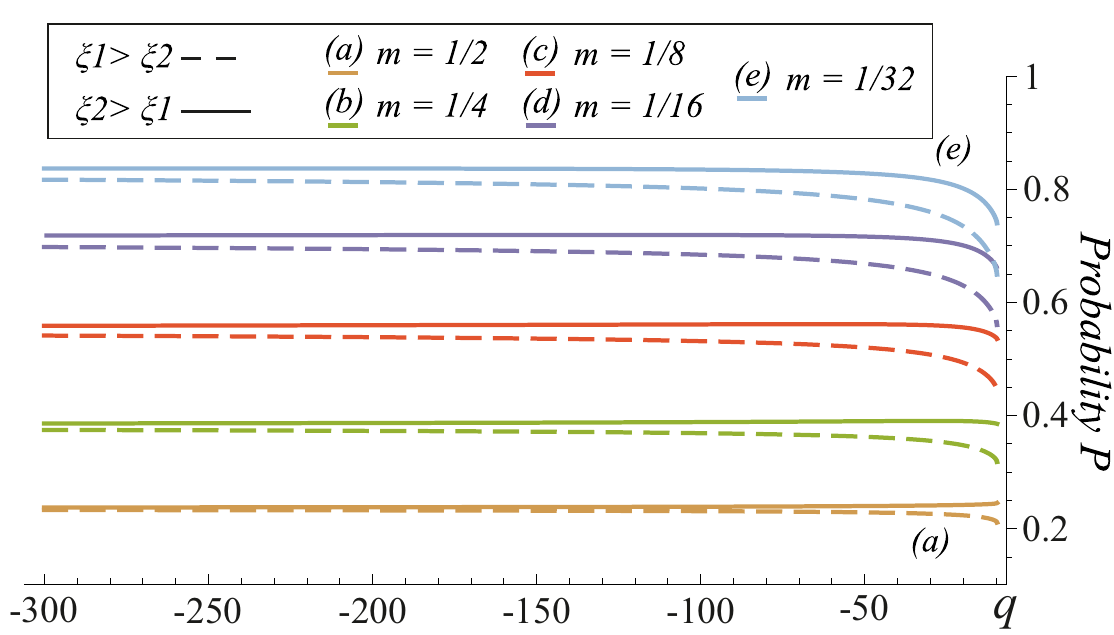}
			\caption{Probability of generating the $\hbox{SWAP}_d^\alpha$ gate as a function of the parameter $q$ for infinite-dimensional systems (limiting case) for $\theta_1-\theta_2=\pi$ and for various entangling parameters $m=\frac{1}{2}\sin(\alpha \pi)$.}
            
            \label{Fig1}
		\end{figure}
		\begin{figure}[h!]
			\includegraphics[width=8.6cm]{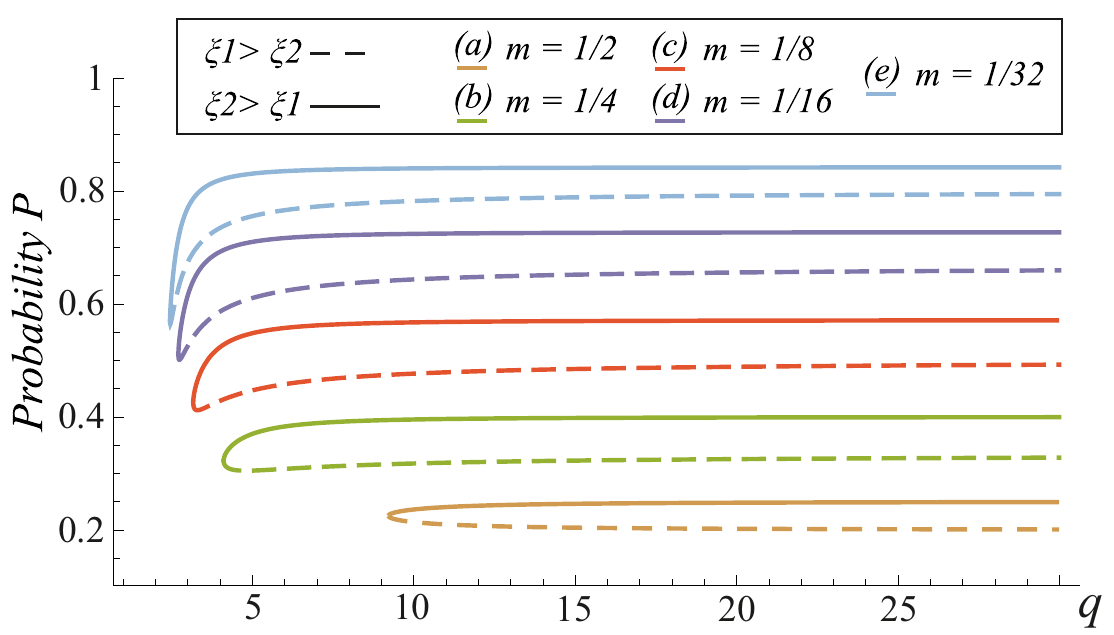}
			\caption{Probability of generating the $\hbox{SWAP}_d^\alpha$ gate as a function of the parameter $q$ for  infinite-dimensional systems (limiting case) for $\theta_1-\theta_2=0$ and for various entangling parameters $m=\frac{1}{2}\sin(\alpha \pi)$.
            }
            \label{Fig2}
		\end{figure}
		
Figures \ref{Fig1} and \ref{Fig2} show the probability curves for the two-qudit transformation with the entanglement parameter $m$. The protocol has four distinct operation regimes: two are shown in \ref{Fig1} as solid and dashed lines, and the other two in\ref{Fig2} using the same line styles. The difference between the rotation angles of the qudits, $\theta_1$ and $\theta_2$, and ratio of the first and second coupling constants, $\xi_1$ and $\xi_2$, determine these regimes. All regimes exhibit asymptotic convergence of the probability to certain limits as $|q|\rightarrow\infty$. The smaller the entanglement parameter $m$, the larger these limits turn out to be.
        
For the regime $\theta_1-\theta_2=\pi$ in Fig. \ref{Fig1}, the solid and dashed curves converge to the same limit which occurs irrespective of the ratio of the coupling constants. The value of this limit depends on the entangling parameter. For $\theta_1-\theta_2=0$ in Fig. \ref{Fig2}, the pattern changes. The two distinct cases, distinguished by the coupling-constant ratio, converge to different limits. As the entangling parameter decreases, the minimum admissible value of the variable $q$ for which the transformation is realized also decreases. The transformation is realized only above this minimum value. Consequently, transformations with lower entangling power not only exhibit higher probabilities but also require smaller values of the coupling constants.
		
Thus, for the probability of generating a unitary transformation at fixed values of the entangling power and the system dimension, there can generally exist three limits to which the probability function may converge. These values are determined by the ratio of the coupling constants $\xi_1 > \xi_2$ or $\xi_2 > \xi_1$ and by the value of $\cos{(\theta_1-\theta_2)} = \pm 1$.
		
It is also of interest to track how the probabilities change for different values of the entangling parameters as the dimension of the logical space varies.
		\begin{figure}\includegraphics[width=8.6cm]{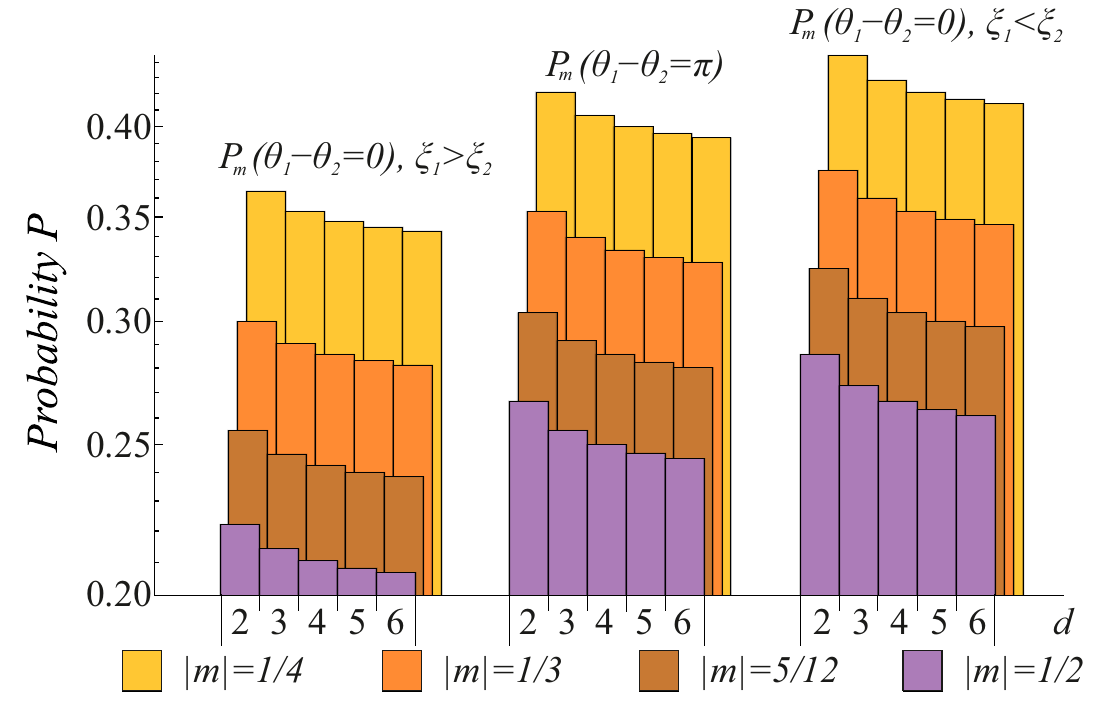}
			\caption{ Asymptotic values of the entanglement generation probability (at $q=\xi_1\xi_2\cos(\theta_1-\theta_2)\rightarrow\infty$) as a function of the entanglement parameter $m$ for various qudit dimensions $d$, shown for three distinct operation regimes of the protocol: for the angle difference $\theta_1-\theta_2=\pi$, irrespective of the ratio of the interaction constants $\xi_1, \xi_2$, the probabilities converge to the same limit; for $\theta_1-\theta_2=0$, there are two probability values, different for $\xi_1>\xi_2$ and $\xi_1<\xi_2$.}
            
            \label{Fig3}
		\end{figure}
		
Figure \ref{Fig3} shows the asymptotic limits of the probability for various operation regimes of the protocol. The regime $\xi_1<\xi_2$ yelds the highest asymptotic probabilities values for all entangling powers. As the entangling power increases, the probabilities decrease for all dimensions of the logical space, as expected. This decrease occurs due to an increase in the probability amplitude of the states with excitation bunching, which invariably accompany the boson entanglement in linear (statistics-preserving) interaction. For larger space dimensions, more bunching possibilities exist in a pair of modes within the same medium. The probabilities decrease almost monotonically for all entanglement parameters and in all regimes as the qudit-space dimension increases, with the maximum occurring for a two-qubit system and the minimum for the formally maximal dimension. The largest leap in the probability upon transitioning to a logical space of a different dimension occurs for the transformation with the maximum entangling power parameter $m=1/2$, which corresponds to the gate $\sqrt{\hbox{SWAP}_d}$. In contrast, the non-entangling $\hbox{SWAP}$ gate will be generated deterministically in a space of any dimension.
		
		\section{Conclusion}
In this work, we have studied the generation of entangling two-qudit operations in the QRQ protocol based on the Faraday interaction between a multimode quantum field carrying OAM and an ensemble of four-level atoms assisted by a strong driving field. We have formulated a method for encoding qudit logical states via atomic and field states with a single excitation in a superposition of OAM-carrying modes. The discreteness of orbital angular momentum allows encoding qudits of arbitrary dimension. It has been shown that the set of physical control parameters of the scheme ensuring the unitarity of the two-qudit transformation matrix does not depend on the logical space dimension. Furthermore, the entangling power calculations demonstrate that the protocol enables the generation of logical $\hbox{SWAP}^\alpha_d$ gates for any value of the power $\alpha$. The conditions that realize a transformation with a given fixed entangling parameter $m=1/2 \sin(\alpha\pi)$ are also independent of the space dimension.

We have also obtained estimates of the two-qudit entangling transformation probability. Since, in the general case, the probability depends on the probability amplitudes of the input qudit state, we derive a lower-bound estimate, i.e., under the assumption that the probability of the vacuum and excitation-bunching terms reach their maximum values. Then, we identify three distinct regimes of the protocol, determined by the difference between the qudit-state rotation angles and by the ratio of the effective coupling constants which can be associated with the interaction times. The operation regimes of the protocol are characterized by different asymptotic behaviors of the probability as the product of the coupling constants formally tends to infinity. The highest  probability values are achieved for a zero angle difference and $\xi_1<\xi_2$. Taking into account that the sum of squares $\xi_1^2+\xi_2^2$ is fixed by the entangling parameter, and the sum of angles is constrained by the unitarity of the transformation, this condition can be formulated as $\xi_1=\sqrt{2},\xi_2\rightarrow\infty,\theta_2=\theta_1=\pm\pi/4$. Then, the probability limit of the maximally entangling transformation $\sqrt{\hbox{SWAP}_d}$ with $m=1/2$ for an arbitrary space dimension (the subscript of the probability indicates the entangling parameter) is given by:
		\BY
		\lim\limits_{\xi_2\rightarrow\infty}P_{1/2}(\xi_1=\sqrt{2},\theta_2=\theta_1=\pm\pi/4)=\frac{d}{4d-1}.
		\EY
Given that the qudit analog of the maximally entangling $\hbox{CNOT}$  operation can be constructed similarly to the qubit case:
		\BY
		\hbox{CNOT}_d=\sqrt{\hbox{SWAP}_d}\cdot D_{s,l}\otimes I_d\cdot\sqrt{\hbox{SWAP}_d},
		\EY
where $D_{s,l}$ is an arbitrary Heisenberg-Weyl operator other than the identity. We can then apply the QRQ protocol twice with an intermediate single-qudit gate to implement two-qudit gate with entangling power equivalent to that of $\hbox{CNOT}_d$. The single-qudit local transformation on the light system can, in principle, be performed deterministically using linear optical elements. Under this assumption, $\hbox{CNOT}_d$ can be implemented in the considered protocol with a probability of at least $1/16$ (for a formally infinite space dimension) and up to $4/49$ for qubits.
		
Thus, we showed that two-qudit entanglement can be realized in the light-atom interaction protocol. This entanglement is essentially realized simultaneously with information storage, since one of the qudits is encoded in a long-lived degree of freedom.

		\appendix
    \section{Estimation of probability amplitudes}
According to Section IIIA, the probability amplitudes of the vacuum and excitation-bunching states depend not only on the values of the physical parameters but also on the qudit state probability amplitudes  $c_i$ and $t_j$. In this appendix, our task is to estimate the maximum values of the factors that depend only on $c_i$ and $ t_j$ in order to determine the worst-case probability of performing the two-qudit operation. We start by finding the maximum of the expression $(1 +\sum\limits_{i\neq j}c_it_jc^*_jt^*_i)$ using the Lagrange multiplier method and, for simplicity, assuming all amplitudes to be real. Let us begin with the following identity:
		\BY
		\sum_{i\neq j}c_it_jc_jt_i = \(\sum_{i}c_it_i\)^2 - \sum_{i}(c_it_i)^2.
		\EY
Given that the probability amplitude of the unitary transformation $U_d$ does not depend on the dimension of the system, we write the Lagrange function $F$ in the following form:
		\BY
		F = \(\sum_{i}c_it_i\)^2 - \sum_{i}(c_it_i)^2 - \gamma_1\sum_{i}c_i^2 -\gamma_2\sum_{i}t_i^2,\;\;\;
		\EY
where $\gamma_i$ are Lagrange multipliers. Equating the partial derivatives $\displaystyle \frac{\partial F}{\partial c_j}=\frac{\partial F}{\partial t_j}=0$, we obtain a condition of the form
		\BY
			c_j^2\gamma_1=t_j^2\gamma_2.
		\EY
We sum this expression on the left-hand and right-hand sides over $j$. This operation yields the equality $\gamma_1 = \gamma_2$. The equality leads to the maximum condition:
	\BY
		|c_j|=|t_j|.
	\EY
Then, the original expression (A1) can be written in the following form:
	\BY
	\sum_{i\neq j}c_it_jc_jt_i =	\sum_{i\neq j}c_i^2c_j^2= 1-\sum_{i}c_i^4.
	\EY
We employ the mean inequality:
	\BY
	\sqrt{\frac{1}{d}\sum_{i}c_i^4}\ge\frac{1}{d}\sum_{i}c_i^2=\frac{1}{d}.
	\EY
Thus, we have:
	\BY
	1-\sum_{i}c_i^4\le1-\frac{1}{d}.
	\EY
The equality holds, for example, when $c_i = t_j = \frac{1}{\sqrt{d}}$. Substituting the obtained conditions into Eq. (A1) results in each term of the sum being equal to $\frac{1}{d^2}$, and the total number of such terms is $d(d-1)$. Consequently, the factor multiplying excitation bunching term becomes:
	\BY
\max\[(1 +\sum\limits_{i\neq j}^{d-1}c_it_jc^*_jt^*_i)\]=	1+\frac{d(d-1)}{d^2}=2-\frac{1}{d}.
	\EY
Similarly, we can find the maximum of the expression for the vacuum term. Repeating the steps described above, it follows that the maximum of the expression $(\sum_i^{d-1}{c_it_i})^2$ is also achieved at $c_i = t_j = \frac{1}{\sqrt{d}}$. In this case, each term of the sum takes the value $\frac{1}{d}$, and there are $d$ such terms in total. This leads to a value equal to unity:
	\BY
		\max\[\sum_{i=0}^{d-1}({c_it_i})^2\]=	1.
	\EY
		
Then the maxima of the probability amplitudes over all possible input states to establish a lower bound on the two-qudit gate implementation probability, can be represented in the following form:
		\BY
		&&|\tau^{est}_d|^2=(2-\frac{1}{d})\(|G_{11} G_{12}|^2+|G_{21} G_{22}|^2\),\\
		&&|\sigma^{est}_d|^2=|\( R_{12}G_{22}+ R_{11}G_{21}\)|^2.
		\EY
        
\section{Constraints on the scheme parameters for performing a unitary operation with a fixed entangling power}
In this appendix, to facilitate further calculations, we switch from the physical parameters of the system to the following set of computational variables:
\BY
&&	\phi=\theta_1+\theta_2,\nn\\
&&q = \xi_1 \xi_2\cos{(\theta_1-\theta_2)},\;\;w=\xi_1 \xi_2\sin{(\theta_1-\theta_2)},\\
&&s=\xi_1^2+\xi_2^2,\notag
\EY
The unnormalized coefficients $\mu$ and $\nu$, defined by relations in Eq. (\ref{munu}), are expressed in terms of the computational parameters as follows:
		\BY
	\mu&=&\frac{1}{8} \left(\left(q^2+w^2\right) \cos (\phi )-(q-4) \sqrt{q^2+w^2}\right)\nn\\
	&&+\left(1-\frac{q}{2}\right) e^{i \phi },\;\;\;\;\;\;\;\;\\
	\nu &=&-\frac{1}{4}e^{i\phi}(s+2q).
	\EY
We factor out the phase $e^{i \phi}$ from both expressions. As discussed in Section II, this allows us to reformulate the unitarity condition of Eq.(\ref{eq:unitar}) in simplified form:
	\BY
	\Re[\mu e^{-i \phi}]=0.
	\EY
This equation is most easily solved for the angle $\phi$, yielding the following set of solutions:
    \BY	
    \phi=\pm\arccos\(\frac{q-4\pm\sqrt{(-16+q (8+ q))}}{2 \sqrt{q^2+w^2}}\).\label{cos_phi}	
	\EY
Thus, there exist four different expressions for the angle $\phi$ for which the two-qudit transformation matrix becomes unitary. For convenience, we will further number these solutions from 1 to 4, considering the case with two "$+$" signs in Eq. (B5) as the first solution, the case with two "$-$" signs as the fourth, and the solutions with "$-+$" and "$+-$" signatures as the second and third, respectively. Note that these solutions, as expected, do not depend on the dimension of the system and were obtained earlier in Ref. \cite{vashukevich2024parallel}.
	
Here we derive expressions for the computational parameters in terms of the entangling power; to this end, we rewrite Eq. (\ref{EnPC}) in an alternative form:
    \BY
    &&\frac{\mu^2\nu^2}{ \(\mu^2 +\nu^2\)^2}=\frac{1}{4}\sin^2(\alpha).
    \EY
Recall from Eq. (B4) $\mu e^{-i \phi}$ must be purely imaginary, and simplify the expressions for $\operatorname{Im}[\mu]$ (here and henceforth we will omit the common phase $e^{-i \phi}$ in the notation for simplicity):
	\BY
        \Im\[\mu\]&=&\(\Re\[\mu\]-\left(1-\frac{q}{2}\right)\)\tan(\phi)
        \\
        \nonumber
        &=&\left(\frac{q}{2}-1\right)\tan(\phi).\;\;\;\;\;\;\;
	\EY
Then, Eq. (B6) takes a simple form by denoting the right-hand side as $m^2$:
		\BY
		\frac{2 (q-2) (2 q+s) \tan (\phi )}{(2 q+s)^2+4 (q-2)^2 \tan ^2(\phi )}=\pm m.
		\EY
Since the obtained values of the angle $\phi$ do not depend explicitly on $s$, we can solve the above equation for $s$:
	\BY
	s=&&\frac{-2 m q+(q-2) \tan (\phi )}{m}\nn\\
	&&\pm\frac{\sqrt{\left(\left(1-4 m^2\right) (q-2)^2 \tan ^2(\phi )\right)}}{-m}.\;\;\;\;
	\EY
Thus, we have eight different sets of solutions $\{s_j(\phi_i), \phi_i\}$ for $ j\in\{1,2\}$ and $i\in\{1,2,3,4\}$, which allow us to calculate the generation probability. 

It is worth mentioning the case when the angle $ \phi= \pi/2+ \pi k, k \in \mathbb{Z}$ and the solutions described by the Eq. (B9) are no longer physically correct. In this case, the Eq. (B6) can be written as
\BY
\Im\[\mu\]&=&(q-4)\sqrt{q^2+w^2}.
\EY
Substituting the expression above into an Eq. (B6), we can get solutions for $s$. In the main text of the work, we will investigate this case in detail.

A minor complication arises when transitioning from the set of computational parameters back to the physical ones, since $\xi_1$ and $\xi_2$ are expressed in terms of $s,q$ and $w$ in a non-unique manner: 
	\BY
	\xi_1 =\frac{\sqrt{
	s \mp \sqrt{-4 q^2 + s^2 - 4 w^2}}}{\sqrt{2}}, \xi_1\lessgtr\xi_2;\\
	\xi_2 =\frac{\sqrt{
		s \pm \sqrt{-4 q^2 + s^2 - 4 w^2}}}{\sqrt{2}}, \xi_1\lessgtr\xi_2,
	\EY
and the eight solutions exhibit branching associated with physically distinct operating conditions of the protocol, namely $\xi_1>\xi_2$ and $\xi_1<\xi_2$.
	
Analysis of the two-qudit transformation probabilities with a fixed entangling power specified by the entangling parameter $m$ shows that the probabilities are maximized when the computational parameter $w=\xi_1\xi_2 \sin(\theta_1-\theta_2)=0$, corresponds to
the values of the rotation angles difference between the atomic and light qudits $\theta_1-\theta_2=\pi k, k\in\mathbb{Z}$. In this case, the transformation probabilities  $P_m(s_j,\phi_i,q)$ with entangling power $m$ for several distinct solutions demonstrate the same dependence on the remaining free parameter $\pm q=\xi_1,\xi_2$:
	\BY
        &&	P_m(\phi_2,q>0)=P_m(\phi_3,q<0)\\\nn
        && = P_{-m}(\phi_1,q>0)=P_{-m}(\phi_4,q<0),\\
        &&	P_{m}(\phi_1,q<0)=	P_{-m}(\phi_2,q<0).
	\EY
In the main text, we discuss the case $P_{m}(\phi_1)$ in greater detail, as it provides the highest probability values.
		
	\end{document}